\documentclass[12pt]{iopart}

\usepackage{graphicx}
\usepackage{color}
\usepackage{bm}
\usepackage{amssymb}

\begin{document}

\title[Ultrafast exciton-polariton scattering towards the Dirac points ]{Ultrafast exciton-polariton scattering towards the Dirac points }

\author{V.~M.~Kovalev$^{1,2}$, I.~G.~Savenko$^{3,4,5}$ and I.~V.~Iorsh$^5$}
\address{$^1$ A.V.Rzhanov Institute of Semiconductor Physics, Siberian Branch of the Russian Academy of Sciences, Novosibirsk, 630090, Russia}
\address{$^2$ Novosibirsk State Technical University, Novosibirsk, 630073, Russia}
\address{$^3$ COMP Centre of Excellence at the Department of Applied Physics, P.O. Box 11000, FI-00076 Aalto, Finland}
\address{$^4$ Center for Theoretical Physics of Complex Systems, Institute for Basic Science, Daejeon, Korea}
\address{$^5$ National Research University of Information Technologies, Mechanics and Optics (ITMO University), Saint-Petersburg 197101, Russia}

\vspace{10pt}
\begin{indented}
\item[]November 2015
\end{indented}

\begin{abstract}
Using the Feynman-Dyson diagram technique, we study nonlinear polariton-polariton scattering in a two-dimensional micropillar-based optical superlattice with hexagonal symmetry. We demonstrate that both the emerging polariton chirality and the loop Feynman diagrams up to infinite order should be strictly accounted for in the evaluation of the self-energy of the system. Further, we explicitly show that in such a design the time of polariton scattering towards the Dirac points can be drastically decreased which can be used, for instance, in engineering novel classes of polariton lasers with substantially reduced thresholds.
\end{abstract}

%
%
%
%
%






\section{Introduction}
Mapping the properties of a fermionic system to a bosonic gas localised in an artificially created complex potential landscape is a powerful tool for probing new condensed matter phenomena. This approach is known as quantum simulation~\cite{QuantSim} attracting vast scientific interest in recent years since it allows to use well-controllable bosonic quantum systems to study far less controlled fermionic properties. Quantum simulation is quite new due to the fact that different bosonic systems, such as ultracold atomic gases~\cite{AtomsSim}, arrays of photonic waveguides~\cite{WaveguidesSim}, and exciton-polariton (later, \textit{polariton}) gases~\cite{CiuttiPolGases}, have just become available experimentally.

The latter system has certain objective advantages over the others due to the mixed light--matter nature of exciton polaritons (EPs). Indeed, on one hand, it became relatively easy and cheap to design and implement arbitrary linear periodic potentials for the photons, utilizing photonic crystals or metamaterial structures, thus particles can be localized and conveniently investigated. On the other hand, due to the excitonic component, the gases of EPs are characterized by strong nonlinear properties resulting from exciton-exciton Kerr-like scattering which represents a superposition of dipole-dipole and exchange interaction. It is important to note, that beside mentioned properties inherited from excitons and photons separately, EPs as hybrid modes with peculiar dispersion, may form quasi-Bose--Einstein condensate (BEC) in the steady state at sufficient background pumping intensities.

Polariton superlattices (later, \textit{lattices}) of different geometry have been recently extensively studied both theoretically and experimentally~\cite{FBAND, DIRAC, SOItheor, SOIexperiment, TI, EdgeStatesExp} with most of attention paid to honeycomb and the Kagome lattices. While at the moment most of the theoretical efforts are dedicated to the linear dispersion features and topological properties of the lattices, it becomes clear that the nonlinear phenomena of such systems should also be substantially affected by the linear spectrum and chirality of polaritons in the hexagonal lattice. In a recent work~\cite{Bogoliubons} it is theoretically shown that chiral edge Bogoliubov states can be excited in a weakly interacting polariton gas in a Kagome lattice potential. Spin-orbit coupling originating from the anisotropy of the hopping coefficient has also been studied both theoretically~\cite{SOItheor} and experimentally~\cite{SOIexperiment}.
Moreover, similar structures have been proven to serve as topological insulators supporting chiral edge states~\cite{TI,EdgeStatesExp}.

It is well known that ultrafast relaxation (the order of femtoseconds) of hot electrons in graphene is dominated by the intraband Coulomb scattering~\cite{BookMalik}. Moreover, it has been shown in~\cite{PRB_Polini} that the intraband electron-electron scattering in graphene is mediated by the collinear processes, which counter-intuitively play significant role in the scattering dynamics, even though they appear along the one-dimensional lines on the dispersion surface and thus, for the first glance, should give zero contribution to the behavior of two-dimensional Dirac fermions~\cite{Coll1,Coll2,Coll3}.

It seems feasible to expect that a bosonic system interacting via a short-range potential (analogously to the screened Coulomb potential for electrons in graphene) and characterized by a linear dispersion should also demonstrate similar ultrafast interaction with induced relaxation to the Dirac points in a certain frequency range.
In this article, we show that the polariton-polariton interaction can be substantially enhanced in the vicinity of the Dirac points. Such a behaviour can be expected if we imply the mentioned quantum simulation mapping between the bosonic and fermionic systems, recalling that the graphene is characterized by ultrafast electron relaxation caused by particle-particle interaction.

Speeding up the polariton scattering to the ground state can be used in various situations~\cite{PolLas0}. One of the straightforward possible applications is a polariton laser. In this device, coherent radiation from the system occurs due to spontaneous emission from a macroscopically occupied single-particle ground state. Since no population inversion is required, such laser is characterized by smaller threshold than conventional stripe edge-emitting or vertical cavity surface emitting lasers. Polariton lasers are usually pumped nonresonantly, optically or via electric current injection~\cite{ElInjPolLas}. The threshold here depends on the ratio of particle lifetime and the complex effective scattering time, -- the time required for the polaritons to reach the ground state from the reservoir of excited states. Using our proposal, it becomes possible to further decrease the scattering time and thus decrease the lasing threshold.


\section{System and model}
\begin{figure}[!t]
\center\includegraphics[width = 0.5\columnwidth]{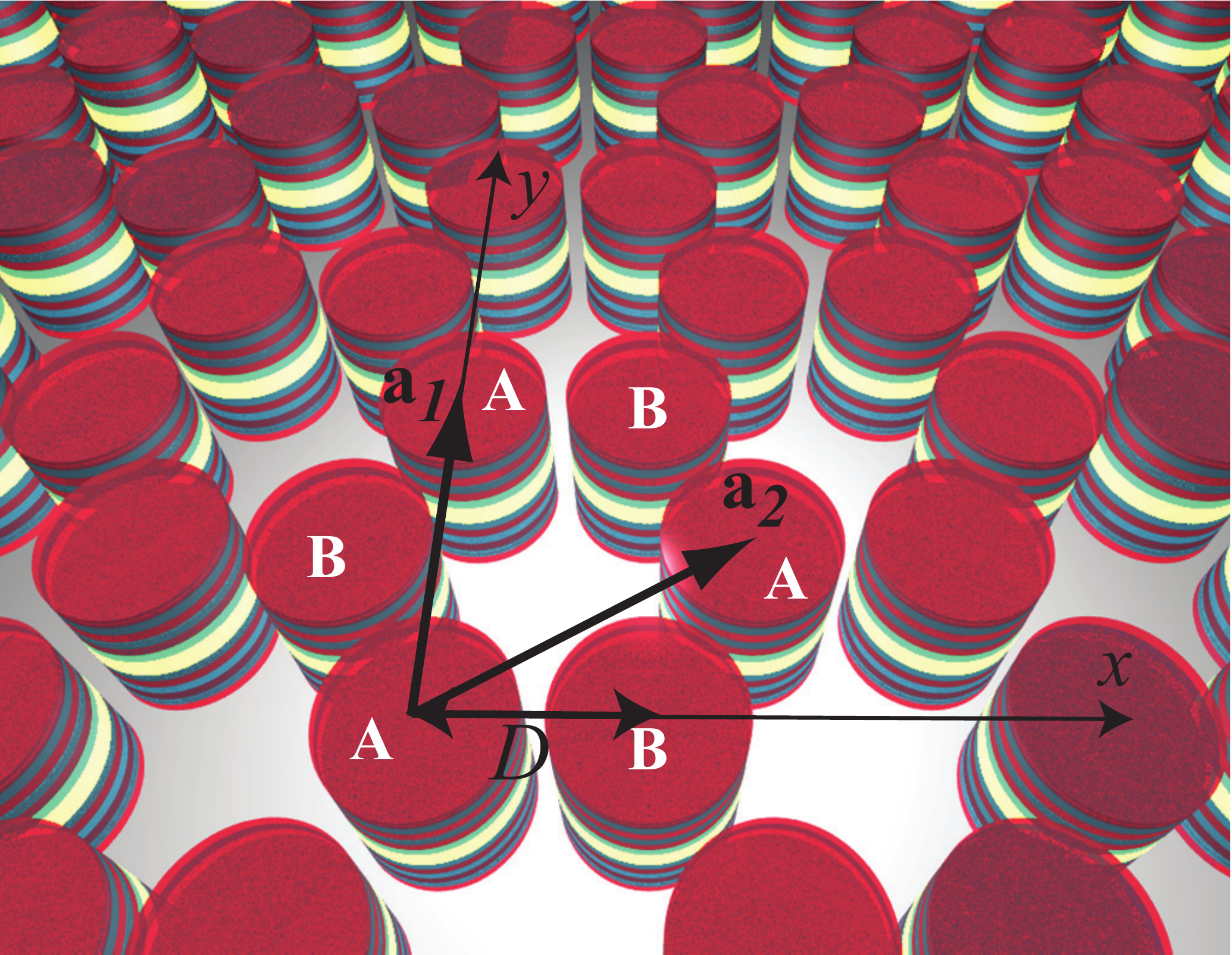}
\caption{Schematic of a hexagonal lattice based on pillar polariton microcavities:  $\mathbf{a}_1$, $\mathbf{a}_2$ here are the translation vectors given by $\mathbf{a}_1=(0,\sqrt{3})D$, $\mathbf{a}_2=(\frac{3}{2},\frac{\sqrt{3}}{2})D$, $x$ and $y$ are the axes labels. Two sublattices are denoted as A and B.}
\label{fig_1}
\end{figure}

We consider a structure with geometry similar to one described in work~\cite{RefKimNJP}, see Fig.~\ref{fig_1}.
It represents a two-dimensional array of micropillar cavities, a design, which can be routinely produced by etching a regular planar semiconductor microcavity. We assume that each micropillar hosts one exciton-polariton mode, and photons are allowed to hop between the neigbouring pillars only.
A unit cell of the lattice consists of two pillars, thus we can denote two sublattices as A and B.
We note that in Ref.~\cite{FBAND} and~\cite{DIRAC}, the authors experimentlly demonstrate existance of the Dirac cones in the dispersion of such structure and polariton BEC in the vicinity of the Dirac point.


The part of the system Hamiltonian corresponding to linear coupling between the subsystems A and B within the tight-binding model reads
\begin{eqnarray}
\label{Ham0}
&\hat{\cal H}
=
\Omega_\mathrm{p}\sum_{i,j,\sigma}\hat{c}^{\dagger}_{\sigma,i,j}\hat{x}_{\sigma,i,j}
-J\displaystyle\sum_{\langle i,j\rangle}\hat{c}^{\dagger}_{A,i,j}\hat{c}_{B,i,j}+\mathrm{h.c.},
\end{eqnarray}
where $i,j,\sigma=A,B$, $\hat{c}_{\sigma,i,j}$ and $\hat{x}_{\sigma,i,j}$ are the annihilation operators of cavity photons and excitons within the sublattice $\sigma$, and $\mathrm{\Omega_\mathrm{p}}$ is exciton-photon coupling strength in each pillar. The neighbouring cavities are coupled via the light components of EPs only with the coupling strength $J$. Here we neglect the spin-orbit coupling originating from the TE-TM splitting of the cavity modes which has been addressed in~\cite{SOItheor}.

Taking the Fourier transform of Eq.~(\ref{Ham0}), we obtain the matrix form of the Hamiltonian in the basis $|C_{A},X_{A},C_B,X_B\rangle$ corresponding to the photon and exciton mode in the $A$ and $B$ sublattice, respectively which takes the form
\begin{eqnarray}
\label{EqHamMatrix}
\hat{\cal H}(\mathbf{k})=
\left(
\begin{array}{cccc}
0 & \mathrm{\Omega}_\mathrm{p} & -JS(\mathbf{k}) & 0 \\ 
\mathrm{\Omega}_\mathrm{p} & 0 & 0 & 0 \\-JS^*(\mathbf{k}) & 0 & 0 & \mathrm{\Omega}_\mathrm{p} \\ 0 & 0 & \mathrm{\Omega}_\mathrm{p} & 0
\end{array}
\right),
\end{eqnarray}
where $S(\mathbf{k})=1+e^{i\mathbf{k}\mathbf{a}_1}+e^{i\mathbf{k}\mathbf{a}_2}$. The eigenenergies of this Hamiltonian can be found by diagonalization the $4\times 4$ matrix in~(\ref{EqHamMatrix}). Similarly to the case of graphene~\cite{Graph_Rev}, the Brilloin zone here contains two inequivalent special  points $\mathbf{K},\mathbf{K}^{\prime}=(\frac{2\pi}{3},\pm\frac{2\pi}{3\sqrt{3}})D$, where $D$ is the distance between $A$ and $B$ sublattices. However, in our case there are two Dirac points with frequencies $\pm \mathrm{\Omega}_p$ located at $\mathbf{K}$ and $\mathbf{K}^{\prime}$.

\begin{figure}[!b]
\centerline{\includegraphics[width = 0.5\columnwidth]{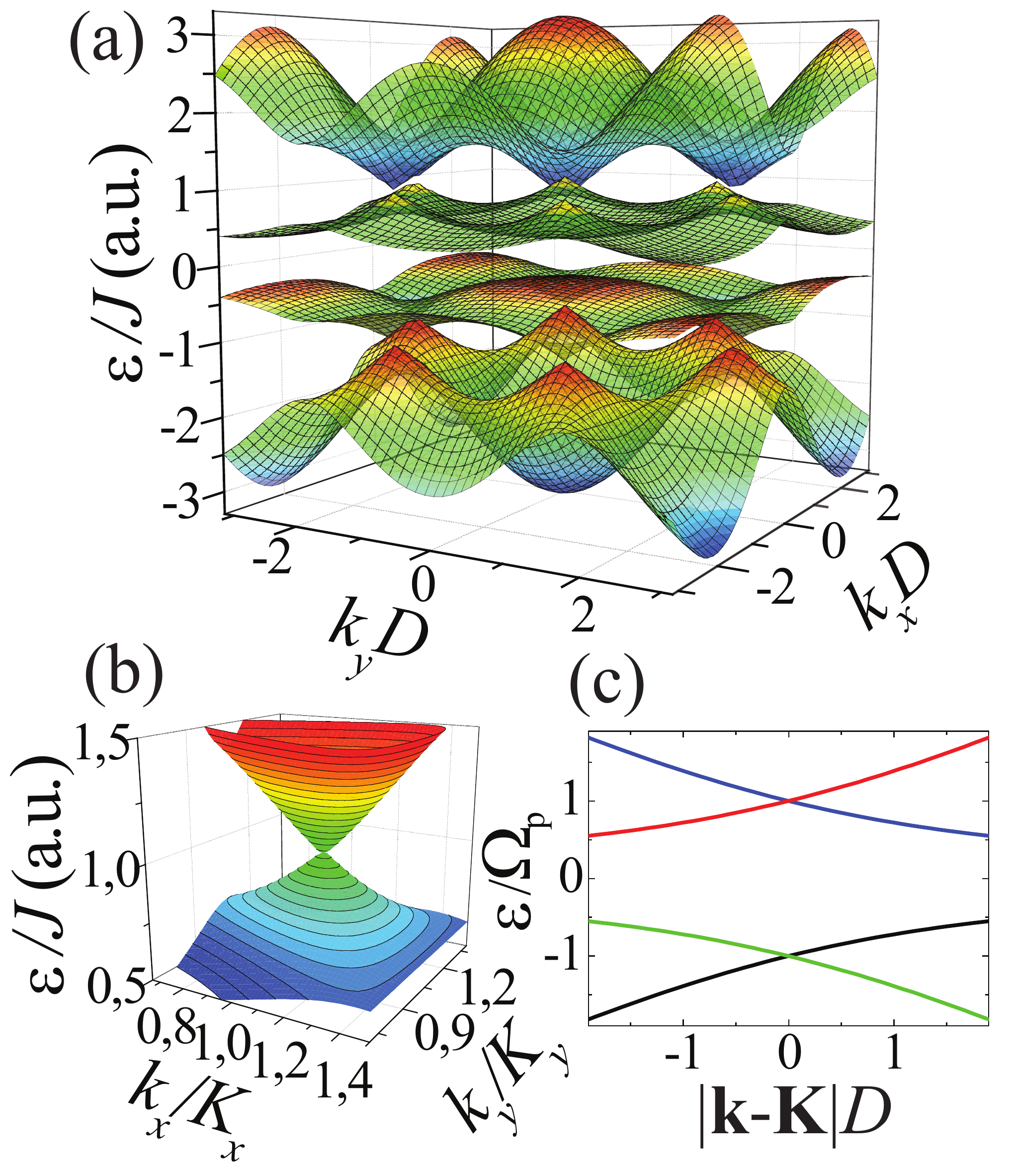}}
\caption{(a) Band structure of the polaritons localised in the hexagonal superlattice.
(b) Dispersion surface of the two upper polaritons in the vicinity of the Dirac point $\mathbf{K}$.
(c) Two-dimensional plot showing four polariton branches in the vicinity of the Dirac point $K$.  }
\label{fig_2}
\end{figure}

The resulting band structure is shown in Fig.~\ref{fig_2}(a). Figure~\ref{fig_2}(b) demonstrates the dispersion in the vicinity of the upper Dirac point $\mathbf{K}$.
Further we expand the Hamiltonian in the vicinity of $\mathbf{K}$ and yield
\begin{eqnarray}
\hat{\cal H}^{(K)}=\lim_{k\rightarrow K}\hat{\cal H}(\mathbf{k})=
\left(
\begin{array}{cccc}
0 & \mathrm{\Omega_p} & v_\textrm{F}ke^{i\phi} & 0 \\ 
\mathrm{\Omega_p} & 0 & 0 & 0 \\ 
v_\textrm{F}ke^{-i\phi} & 0 & 0 & \Omega_\mathrm{p} \\ 
0 & 0 & \Omega_\mathrm{p} & 0
\end{array} 
\right),
\label{Ham2}
\end{eqnarray}
where $v_\textrm{F}=(3JD)/2$ is the Fermi velocity of the photonic mode, and $\phi=\arctan{(k_y/k_x)}$. We then apply the following unitary basis transformation
\begin{eqnarray}
&L_{A}=\frac{C_A-X_A}{\sqrt{2}},\quad U_{A}=\frac{C_A+X_A}{\sqrt{2}}; \nonumber \\
&L_{B}=\frac{C_B-X_B}{\sqrt{2}}, \quad U_{B}=\frac{C_B+X_B}{\sqrt{2}}, \label{transform}
\end{eqnarray}
which allows us to rewrite the Hamiltonian~(\ref{Ham2}) as
\begin{eqnarray}
\hat{\cal H}^{(K)}=
\frac{1}{2}
\left(
\begin{array}{cccc}
-2{\Omega}_{p} & v_\textrm{F}ke^{\mathrm{i}\phi} & 0 & v_\textrm{F}ke^{\mathrm{i}\phi} \\ 
v_\textrm{F}ke^{-\mathrm{i}\phi} & -2\mathrm{\Omega}_\mathrm{p} & v_\textrm{F}ke^{-\mathrm{i}\phi} & 0 \\ 
0 & v_\textrm{F}ke^{\mathrm{i}\phi}
& 2\mathrm{\Omega}_\mathrm{p} & v_\textrm{F}ke^{\mathrm{i}\phi} \\ 
v_\textrm{F}ke^{-\mathrm{i}\phi} & 0 & v_\textrm{F}ke^{-\mathrm{i}\phi} & 2\mathrm{\Omega}_\mathrm{p} \\
\end{array}
 \right).
\end{eqnarray}
We see that there are two chiral polariton branches in the vicinity of energy $\mathrm{\Omega}_\textrm{p}$ and two in the vicinity of $-\mathrm{\Omega}_\textrm{p}$.
It is also clear from the matrix above, that these modes are coupled. However, 
if we put ourself in the vicinity of the Dirac points, such that $v_\textrm{F}k\ll 2\mathrm{\Omega}_\textrm{p}$, 
we can omit the coupling of upper and lower Dirac cones since the energy correction due to these 
couplings is proportional to $(v_\textrm{F}k)^2/(2\mathrm{\Omega_p})^2\ll1$. 
Then the Hamiltonian takes a block-diagonal form and splits on two $2\times2$ matrices, 
and we can describe each of the Dirac cones separately. 
Hence the $2\times2$ Hamiltonian of the two low-polariton chiral branches is given by
\begin{eqnarray}
\hat{\cal H}^{(2\times 2)}=
\left(
\begin{array}{cc}
-\mathrm{\Omega}_\mathrm{p} & \frac{1}{2} v_\textrm{F}ke^{\mathrm{i}\phi} \\ \frac{1}{2} v_\textrm{F}ke^{-\mathrm{i}\phi} & -\mathrm{\Omega}_\mathrm{p} 
\end{array}
\right)
 \end{eqnarray}
 with eigenenergies
\begin{eqnarray}
 \varepsilon_{\textbf{k}}=-{\Omega}_\mathrm{p}\pm \frac{1}{2}v_\textrm{F}k.
\label{PolDisp}
\end{eqnarray}
%



\section{Nonlinear scattering}
In order to account for the polariton-polariton scattering, we consider another term of the Hamiltonian,
\begin{equation}
 \hat{\cal H}_\textrm{int}=g\int d\textbf{r}\left[\hat\Psi^{+}(\textbf{r})\hat\Psi(\textbf{r})\right]^2,
 \label{IntHam}
\end{equation}
where $\hat\Psi(\textbf{r})$ is a polariton field operator. Thus we investigate contact--like polariton--polariton interaction. For simplicity, we assume that the Hopfield coefficients for excitonic and photonic fractions are the same, $\chi=1/\sqrt{2}$, hidden in the coupling strength $g$.
Our main goal is to find the analytic formula for the polariton scattering time (inverse scattering rate). 

Let us assume that the system is homogeneous at the initial moment of time with zero number of particles. Then it is excited with energy of external pump above the Dirac cone. A polariton with initial momentum $\mathbf{q}$ can be scattered towards the state $\mathbf{p}$ with the transferred momentum $\mathbf{k}$. The momentum and energy conservation laws imply
\begin{equation}
q+p=\sqrt{q^2+k^2-2qk\cos\alpha}\pm\sqrt{p^2+k^2+2pk\cos\beta}, \label{Eq:EnConserv}
\end{equation}
where $\alpha$, $\beta$ are the angles between vectors $\mathbf{k}$ and $\mathbf{q}$, $\mathbf{k}$ and $\mathbf{p}$, respectively. The plus sign in Eq.~(\ref{Eq:EnConserv}) corresponds to the intraband scattering, and the minus sign to the interband one. It can be easily shown that in the case of interband process, only the collinear scattering $\alpha=\beta=0$ is allowed. This means that in the case of broad in $\mathbf{k}$ initial distribution, the rate of the interband processes is negligible as compared to the interband scattering rate. Even the account of finite polariton lifetime does not lift this reduciton. Therefore, in what follows, we assume that EPs occupy only the positive energy states.
 
\begin{figure}[!t]
\centerline{\includegraphics[width = 0.8\columnwidth]{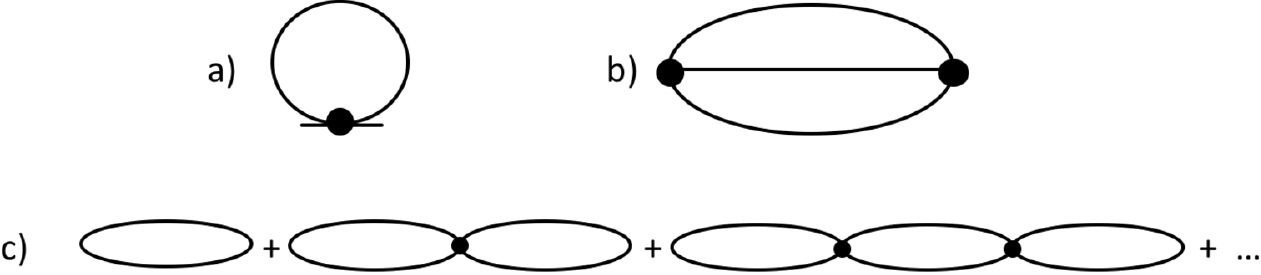}}
\caption{Feynman diagram of the polariton--polariton scattering process.}
\label{Fig3}
\end{figure}

We now use the diagrammatic approach in order to find the polariton scattering lifetime. It is well known, that the Feynman diagram of the first order in the polariton interaction constant $g$ [see Fig.~\ref{Fig3}(a)] does not contribute to the scattering time.
In the second order in $g$, the self-energy operator is given by the diagram depicted in Fig.~\ref{Fig3}(b),
\begin{equation}
\label{SelfEn}
\Sigma(i\omega_n;\textbf{q})=g^2T\sum_{\omega_m,\textbf{k}}\Pi(i\omega_m;\textbf{k})G(i\omega_{mn};\textbf{q+k})F(\textbf{q},\textbf{k}),
\end{equation}
where $\omega_{mn}=\omega_m+\omega_n=2\pi (m+n)/(k_BT)$ are bosonic frequencies, $G^{-1}(i\omega_n;\textbf{k})=i\omega_n-\varepsilon_{\textbf{k}}$ is a Green function of the upper-branch polaritons, and $F(\textbf{q},\textbf{k})=\frac{1}{2}[1+\cos(\varphi_{\textbf{q}}-\varphi_{\textbf{q+k}})]$ is their chirality form-factor. 
Note, here and below we measure the energy of the polariton upper branch from $-{\Omega}_\mathrm{p}$ and assume $\varepsilon_{\textbf{k}}=v_0k$, where we denote $v_0=v_\textrm{F}/2$ for convenience. In this case we will obtain the Bosonic distribution with the chemical potential $\mu$ connected with the total polariton density $n_\textrm{pol}$.

%

In the lowest order in $g$, the function $\Pi(i\omega_n;\textbf{k})$ is given by the single-bubble diagram,
\begin{equation}
\Pi(i\omega_n;\textbf{k})=\sum_{\textbf{p}}F(\textbf{p},\textbf{k})\frac{f(\varepsilon_{\textbf{p+k}})-f(\varepsilon_{\textbf{p}})}{i\omega_n+\varepsilon_{\textbf{p}}-\varepsilon_{\textbf{p+k}}},
\label{singlebubble}
\end{equation}
where $f(\omega)=\{\exp{[(\omega-\mu)/T]}-1\}^{-1}$ is the Bose distribution, $\mu$ is the chemical potential. It can be easily demonstrated that the direct calculation via the formula (\ref{singlebubble}) results in a nonphysical behavior of the scattering time at low polariton energies. To find the correct way to go, we take into account all the bubble diagrams, as shown in Fig.~\ref{Fig3}c, and yield an analytic formula for the retarded self-energy in the form
\begin{eqnarray}
\label{selfretard}
\Sigma^R(\epsilon,\textbf{q})&=&-\frac{g^2}{\pi}\sum_{\textbf{k}}F(\textbf{q},\textbf{k})\\
\nonumber
&&\times\int d\omega\frac{f(\varepsilon_{\textbf{q+k}})-f(\omega)}{\epsilon+\omega-\varepsilon_{\textbf{q+k}}+i\delta}\textrm{Im}\,P^R(\omega,\textbf{k}),
 \end{eqnarray}
where $P^R(\omega,\textbf{k})$ is the result of summation of the bubble diagrams by the Dyson equation,
\begin{equation}
P^R(\omega,\textbf{k})=\frac{\Pi^R(\omega,\textbf{k})}{1-g\Pi^R(\omega,\textbf{k})},
\label{bubbleseries}
\end{equation}
where $\Pi^{R}(\omega;\textbf{k})=\Pi(\omega+i\delta;\textbf{k})$.
In the limit of low polariton energy and momenta, $\epsilon\rightarrow 0$, $\textbf{q}\rightarrow 0$, the scattering time is given by the imaginary part of the polariton self-energy: $\tau^{-1}\sim\,\textrm{Im}\,\Sigma^R(\epsilon,\textbf{q})$.
The polarization operator, representing a single-bubble diagram, at small $|\textbf{k}|$ can be written as
\begin{equation}
\Pi^R(\omega,\textbf{k})=v_0\sum_{\textbf{p}}\frac{\partial f(\varepsilon_{\textbf{p}})}{\partial\varepsilon_{\textbf{p}}}\frac{\textbf{e}_{\textbf{p}}\textbf{k}}{\omega-v_0\textbf{e}_{\textbf{p}}\textbf{k}+i\delta},
\label{PolOper1}
\end{equation}
where $\textbf{e}_{\textbf{p}}=\partial \varepsilon_{\textbf{p}}/(v_0\partial \textbf{p})$ is a unity vector along the direction of $\textbf{p}$.

Further we evaluate the integrals and find
\begin{eqnarray}
\textrm{Re}\,\Pi^R(\omega,\textbf{k})&=&\Pi_0\left(1-\frac{|\omega|}{\sqrt{\omega^2-v_0^2k^2}}\theta[\omega^2-v_0^2k^2]\right)
\\
\textrm{Im}\,\Pi^R(\omega,\textbf{k})&=&\Pi_0\frac{\omega}{\sqrt{v_0^2k^2-\omega^2}}\theta[v_0^2k^2-\omega^2],
\label{ImRePolOper}
 \end{eqnarray}
where 
\begin{eqnarray}
\label{EqPi0}
\Pi_0=-\int_0^\infty \frac{pdp}{2\pi}\frac{\partial f}{\partial\varepsilon_{\textbf{p}}}=\frac{T}{2\pi v_0^2}\ln\left[\frac{1}{1-e^{\mu/T}}\right].
\end{eqnarray}
%


\section{Results and discussion}
If we put $\textbf{q}=0$ in Eq.~(\ref{selfretard}) and expand the Bose distribution functions near $\epsilon= 0$, we find
\begin{equation}
\textrm{Im}\,\Sigma^R(\epsilon)=g^2\epsilon\sum_{\textbf{k}}\frac{\partial f}{\partial\varepsilon_{\textbf{k}}}\textrm{Im}\,P^R(\varepsilon_{\textbf{k}}-\epsilon,\textbf{k}).
\label{ImSelfEn2}
 \end{equation}
Substituting Eq.~(\ref{PolOper1}) and~(\ref{ImRePolOper}) in Eq.~(\ref{ImSelfEn2}), we find that the imaginary part of the polariton self-energy reads
\begin{eqnarray}
\textrm{Im}\,\Sigma^R(\epsilon)&=&-\frac{\epsilon^2}{T}{\cal I}\left(\frac{\epsilon}{2T}\right),~~~\textrm{where}\\
\nonumber
{\cal I}(z)&=&\frac{1}{\pi\ln\left[\frac{1}{1-e^{\mu/T}}\right]}\int\limits_1^\infty \frac{dx}{\sqrt{x-1}}\frac{1}{e^{xz-\mu/T}-1}.
\label{ImSelfEn3}
\end{eqnarray}
At small energies $\epsilon\ll T$, when $z\ll 1$, the integral ${\cal I}(z)$ can be approximated as
\begin{eqnarray}
{\cal I}(z)\sim\int\limits_1^{1/z} \frac{dx}{\sqrt{x-1}}\approx 2/\sqrt{z},
\label{ImSelfEn4}
\end{eqnarray}
and the polariton scattering time acquires the following energy dependence:
\begin{eqnarray}
\frac{1}{\tau}\sim \epsilon\sqrt{\frac{\epsilon}{T}}.
\end{eqnarray}
This dependence is the central result of our work. 
\begin{figure}[!t]
\centerline{\includegraphics[width = 0.5\columnwidth]{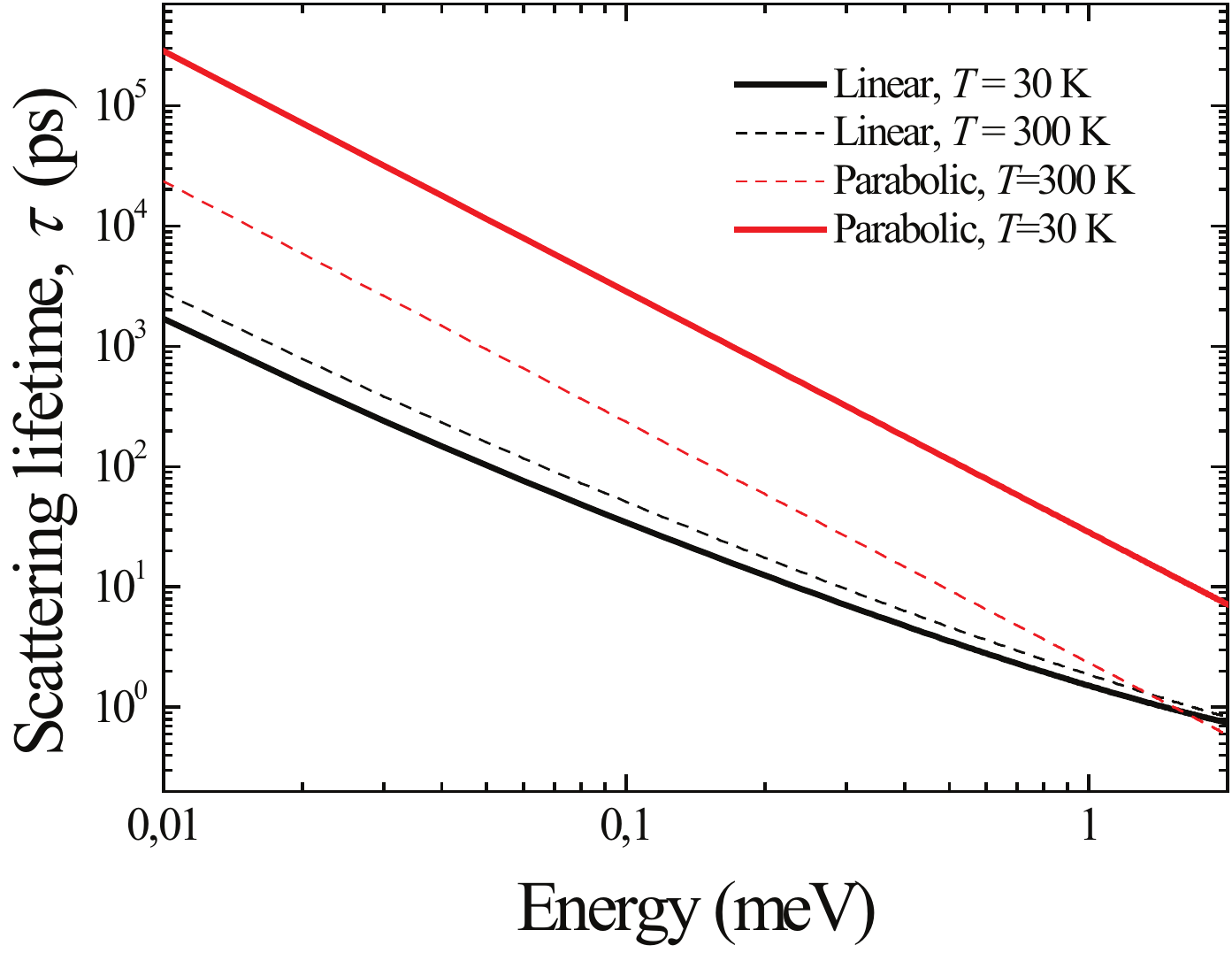}}
\caption{Plot of the scattering lifetime vs the quasiparticle energy for the case of linear and parabolic spectrum and two different temperatures. System parameters are discussed in the text.}
\label{fig_4}
\end{figure}
If we compare our case of Dirac-like dispersion (at small $\mathbf{k}$) with the case of usual parabolic dispersion of EPs at $|\mathbf{k}|\rightarrow 0$, we immediately find that in the latter case $1/\tau\sim \epsilon^2/T$. That is, at $\epsilon\rightarrow 0$, $\epsilon^{3/2}$ decreases slower than $\epsilon^{2}$, and hence linear dispersion is more profitable from the point of view of efficient scattering towards the ground state.
We illustrate it by plotting the scattering lifetime in the vicinity of the ground state for two different temperatures in the case of linear and parabolic spectrum at Fig.~\ref{fig_1}. The chemical potential is $-1$ meV, the effective mass for the parabolic spectrum is set to $10^{-35}$ kg, the exciton binding energy  - $30$ meV, Bohr radius - $20$ nm, and Fermi velocity for the linear case - $3\times10^6$ m/s. We can see that for both cryogenic and room temperatures, the scattering lifetimes are at least one order of magnitude smaller than for the parabolic case, which corresponds to far faster relaxation to the ground state.

It should be additionally stressed, that EPs remain good quasiparticles in the vicinity of the Dirac point, $\epsilon=0$, since~\cite{Bruuc}

\begin{eqnarray}
\frac{1}{\epsilon\tau}\sim\sqrt{\frac{\epsilon}{T}}\ll1.
\end{eqnarray}
%


\section{Conclusions}
In this article, we used the Feynman-Dyson diagram approach to calculate polariton-polariton interaction self-energy in structures with linear Dirac-like dispersion. We explicitely demostrated that the polariton scattering time caused by the nonlinear interaction between particles is dramatically decreased in comparison with conventional two-dimensional planar structures with parabolic dispersion of polaritons. It happens since the scattering lifetime of polaritons is proportional to the power $-3/2$ of the energy rather than power $-2$ which becomes significant at small energies. The effect which we considered can, for instance, be used to construct polariton lasers with significantly reduced thresholds.

\ack{The authors would like to acknowledge the financial support from Grant RFBR 15-02-08949; the Government of the Russian Federation, Grant No. 074-U01; the Dynasty Foundation; and Academy of Finland through its Centre of
Excellence Program under Grant No. 251748 (COMP). }


\end{document}